# Formation of in-volume nanogratings with sub-100 nm periods in glass by femtosecond laser irradiation

Yang Liao[1], Wangjun Pan[1], Yun Cui[2], Lingling Qiao[1], Yves Bellouard[3], Koji Sugioka[4], and Ya Cheng[1,*]

[1]*State Key Laboratory of High Field Laser Physics, Shanghai Institute of Optics and Fine Mechanics, Chinese Academy of Sciences, Shanghai 201800, China*
[2]*Key Laboratory of Materials for High Power Laser, Shanghai Institute of Optics and Fine Mechanics, Chinese Academy of Sciences, Shanghai 201800, China*
[3]*Galatea Lab, STI/IMT, Ecole Polytechnique Fédérale de Lausanne (EPFL), Rue de la Maladière 71b, CH - 2002 Neuchâtel, Switzerland*
[4]*RIKEN-SIOM Joint Research Unit, RIKEN - Advanced Science Institute, Center for Advanced Photonics, Hirosawa 2-1, Wako, Saitama 351-0198, Japan*
*\*Corresponding author: ya.cheng@siom.ac.cn*

We present direct experimental observation of the morphological evolution during the formation of nanogratings with sub-100-nm periods with the increasing number of pulses. Theoretical simulation shows that the constructive interference of the scattering light from original nanoplanes will create an intensity maximum located between the two adjacent nanoplanes, resulting in shortening of the nanograting period by half. The proposed mechanism enables explaining the formation of nanogratings with periods beyond that predicted by the nanoplasmonic model.

Nowadays, femtosecond lasers have become a reliable tool for fabricating microstructures with three-dimensional (3D) geometries in various transparent materials [1]. This is achieved by the unique characteristics of ultrashort pulse widths and extremely high peak intensities of femtosecond laser pulses, which allow efficient confinement of nonlinear absorption of light within the focal volume and dramatic suppression of heat diffusion. Moreover, taking the advantage of a threshold effect in femtosecond laser processing has successfully improved the fabrication resolution far beyond the diffraction limit in both two-photon polymerization and surface ablation [2,3]. Further combination of the threshold effect and a unique phenomenon in the femtosecond laser interaction with glass, namely, formation of self-organized in-volume nanogratings, has enabled fabrication of nanofluidic channels in glass with a width of only ~40 nm [4]. Interestingly, despite its successful applications in a variety of fields such as nanophotonics and nanofluidics, the physical mechanism behind the femtosecond-laser-induced nanogratings has been not fully understood and is still under intensive investigations [5-8].

One difficulty in investigating the formation mechanism of the nanogratings buried in glass is owing to lack of an efficient and non-invasive approach for observing evolution of the nanostructures under the irradiation of femtosecond laser pulses until a stable nanograting structure has been produced. One method consists in observing nanogratings using atomic force scanning thermal microscopy [9,10]. Modified regions have a lower thermal conductivity than unmodified ones. Although this method is applicable to plain silica, it has a typical resolution of 100 nm and therefore, does not allow for lower scale observations. Recently, we overcome this difficulty by focusing femtosecond laser pulses in a porous silica glass immersed in water, in which nanogratings consisting of an array of hollow nanoplanes can be directly produced without chemical wet etching [11]. For investigating the nanograting formation, this is non-trivial because the chemical etching usually used for revealing the subtle modifications in typical glass induced by femtosecond laser irradiation can often destroy the nano-scale features in the laser modified zone. In contrast, direct observation of structural evolution in the porous glass without the chemical etching suggests that surface plasma waves excited at the interface plays a significant role to initiate the formation of nanogratings. Furthermore, considering the type of nanostructures observed [11], we can assume than the porous silica immersed in water also offers a valid model, comparable to plain silica, for observing the formation of self-organized nanoplanes.

An unresolved inconsistency between the surface plasma model and the previous experimental observation of nanograting formation lies in the fact that the period of nanogratings decreases with the increasing number of femtosecond laser pulses, and saturates at the critical point, as previously reported by S. Richter *et al.* [12] in plain silica. The shortest period of nanogratings can be significantly less than that predicted by the surface plasma wave or nanoplasmonic models [11,13]. In the writing of the nanogratings in porous silica glass, the same phenomenon can also be observed when a sufficient number of femtosecond laser pulses at the intensities below the single-shot ablation threshold are deposited in the vicinity of focus. As shown in Fig. 1, a grating with a period as narrow as ~70-80 nm can be produced. Remarkably, the grating period is less than $\lambda/10$ (here, the center wavelength of our writing laser is 800 nm), which is significantly narrower than the period of ~$\lambda/2n$ predicted by the nanoplasmonic model [13], and the minimum period of ~97 nm predicted by the interface plasma wave model [11]. The writing scheme used for producing such grating is illustrated in Fig. 2(a) with all the irradiation conditions directly provided in the caption of Fig. 1.

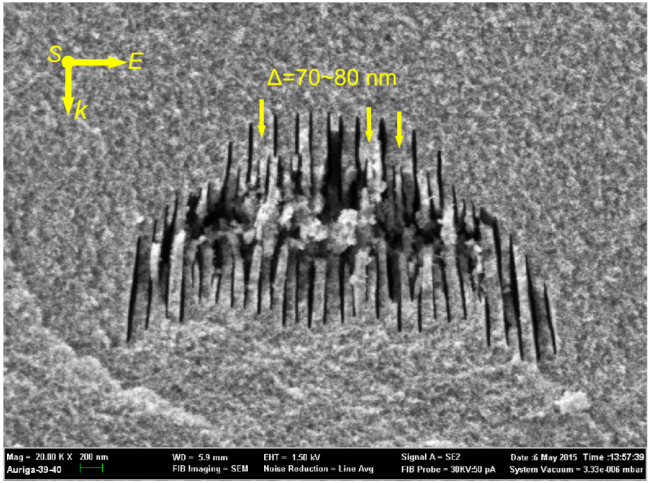

Fig. 1. Cross sectional morphology of short-period nanograting written in porous glass. The pulse energy of femtosecond laser was ~230 nJ after the laser pass through a slit with a width of 640 μm, which was slightly less than the single-shot damage threshold; and the sample translation velocity was set at 20 μm/s, corresponding to that each modified spot was exposed to approximately 6000 pulses. The laser incident direction (**k**), polarization direction (**E**), and the writing direction (**S**) are indicated in the figure.

To fully understand the formation mechanism of the nanograting shown in Fig. 1, in this Letter, we directly visualize the morphological evolution in the porous glass with the increasing number of pulses in the low pulse energy and high pulse number regime. Based on our experimental observation and theoretical modeling, we propose that the constructive interference of the scattering light from two adjacent nanoplanes will create an intensity maximum located at the middle of the two nanostructures. The nanoscale ablation induced by such interference pattern leads to rapid reduction of the period of the nanograting with a high number of irradiation pulses.

In our experiment, high-silicate porous glass samples were used as the substrates, which were produced by removing the borate phase from phase-separated alkali-borosilicate glass in hot acid solution [14]. The composition of the porous glass is approximately 95.5$SiO_2$-4$B_2O_3$-0.5$Na_2O$ (wt %). The pores with a mean size of ~10 nm are distributed uniformly in the glass and occupy 40% volume of the glass. To induce the nanograting structures inside porous glass, a high-repetition regeneratively amplified Ti:sapphire laser (Coherent, Inc., RegA 9000) with a pulse duration of ~100 fs, a central wavelength of 800 nm and a repetition rate of 250 kHz was used.

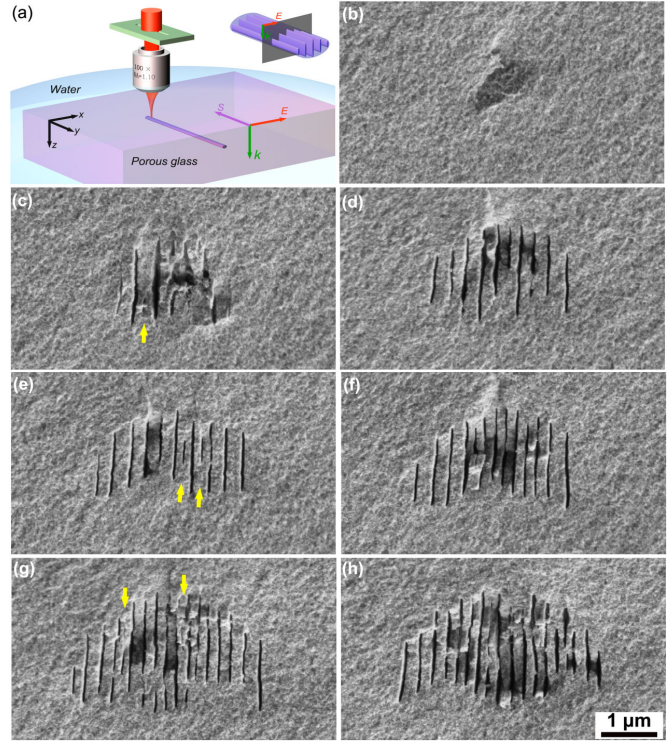

Fig. 2. (a) Experimental setup of direct writing of nanograting in porous glass. (b-h) Cross sectional morphologies of nanogratings written at various scan velocities. Scan velocities in (b-h) are 650, 600, 450, 250, 150, 10, 5 μm/s, and the corresponding effective pulse numbers are 185, 200, 267, 480, 800, 12000, and 24000, respectively.

Figure 2(a) illustrates the experimental setup of femtosecond laser direct writing in the porous glass. A 5×5×2 mm porous glass sample was fixed in a petri dish filled with distilled water, and a long-working-distance water-immersion objective (Olympus, N.A. = 1.10) was employed for focusing the beam into the sample at a depth of ~170 μm below the sample surface. To achieve more homogeneous optical field distribution in the transverse direction and mitigate the influence of nonlinear self-focusing in the propagation direction, a narrow slit was placed above the objective lens, resulting in an expansion of the focal spot in the transverse direction due to the diffraction effect of the slit [15]. Here, to facilitate discussion, we define an artificial parameter which is called effective pulse number N. The effective pulse number is defined as a measure of the laser pulses deposited in the focal area, which can be quantitatively expressed as $N=d \times R/v$, wherein $d$ is the $1/e^2$ width of the focal spot in the scan direction, $v$ is the scan speed, and $R$ is the repetition rate of laser pulse. Based on the parameters of the laser beam and the objective lens, $d$ is calculated to be ~0.48 μm. Thus, by changing the scan speed ($v$), we can investigate the dependence of the period of the nanogratings on the effective pulse number, or in other words, to the energy deposited. To characterize the morphologies of the embedded nanogratings, the fabricated samples were cleaved along the plane perpendicular to the writing direction to access the cross sections of laser written trace. The revealed nanograting structures were directly characterized using an SEM (Zeiss Auriga 40). Neither chemical etching nor annealing was used in the entire experiments.

Figures 2(b-h) present the different morphologies of the structures observed in the porous glass at different numbers of laser pulses. To obtain the results, the slit width was chosen to be 400 μm for producing a more homogenous optical field in the transverse direction (i.e., perpendicular to the laser scan direction). Throughout the experiment, the pulse energy of femtosecond laser was maintained at ~270 nJ after the laser beam passed through the slit, and the number of irradiation pulses was gradually increased by reducing the scan speed. In the beginning, no periodic structure could be observed at the focus at an effective laser pulse number of ~185, as shown in Fig. 2(b). When the effective laser pulse number was increased to ~200, periodically distributed nanovoids were first formed at the interface between the regions modified and unmodified by the laser irradiation, and then the nanovoids can grow into thin narrow planes due to the local field enhancement which causes asymmetric nano-ablation preferentially occurring in the vertical direction, as shown in Fig. 2(c), which is consistent with our previous observation [11]. An interesting feature in Fig. 2(c) is that between one of the two adjacent nanoplanes in the originally formed nanograting, a new nanoplane can be observed as indicated by the yellow arrow. This feature becomes more and more prominent when the number of pulses is increasing, leading to a rapid reduction of the grating period, as shown in Figs. (d-g). Finally, at a pulse number of 12000, the reduction of the grating period saturates, as shown in Figs. 2(g-h). At this point, the grating period has been reduced almost by half as compared to the original grating period shown in Fig. 2(c). Nevertheless, with a large number of pulses, thermal effects due to heat accumulation start to affect the morphology of the formed nanograting, leading to distortion and merge of some nanoplane due to the shockwave generation and melting, as evidenced by Fig. 2(h).

To understand the behavior revealed in Fig. 2, we simulate the optical field intensity distribution of the scattering light from the nanogratings using 3D finite-difference time-domain (FDTD) method [16,17]. First, we assume there are only two nanoplanes separated by ~400 nm, as illustrated in Fig. 3(a). Under the irradiation of the femtosecond laser pulses, a nanoplasma sheet can be produced in each nanoplane, because the nanoplanes are filled with water which has a lower photoionization potential than that of the fused silica [18]. The laser beam used in our simulation is assumed to be linearly polarized with a wavelength centered at 800 nm and a spectral width (FWHM) of ~30 nm. Using Drude model, the relative dielectric constant in the laser field including the effect of plasma density ($N_e$) can be derived as below [19],

$$\varepsilon = \varepsilon_m - \frac{N_e e^2}{\varepsilon_0 m^* m} \frac{1}{(\omega^2 + i\omega/\tau)}, \qquad (1)$$

where $\omega$ is the incident light frequency in vacuum, $\tau$ is the Drude damping time of free electrons, and $\varepsilon_0$, $m$, $m^*$ and $e$ are the relative dielectric constant in vacuum, the electron mass, optical effective mass of carriers and electron charge, respectively. We assume that the nanoplasma has a plasma density ($N_e$) of $5 \times 10^{20}$ cm$^{-3}$. Such plasma density is chosen to best reproduce the experimental observations because direct measurement of the plasma density in the volume of glass is difficult. With $\varepsilon_0$ = 2.1025, $m^*$ = 0.2 [20], and $\tau$ = 10 fs [17], the optical constant of nanoplasma is calculated to be $n$ = 0.81945 + 0.03712$i$.

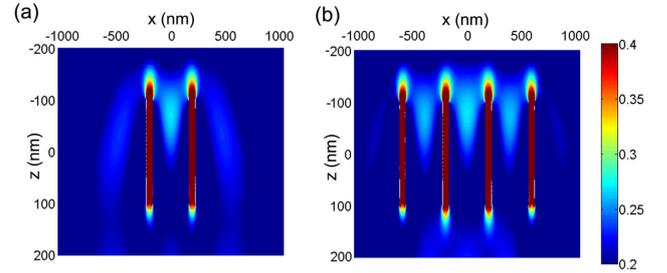

Fig. 3. Distributions of light intensity in XZ plane near (a) two nanoplasma sheets (50 nm × 700 nm × 200nm) separated by an interval of 400 nm, and (b) an array of four nanoplasma sheets (50 nm × 700 nm × 200nm) with an period of 400 nm.

As shown by the simulation result in Fig. 3(a), an intensity maximum appears between the two nanoplasma sheets, which is caused by the coherent superposition of the scattering waves from the nanoplasma sheets. Such a local field enhancement between the two nanoplanes will induce nano-scale ablation and subsequently, the formation of a new "son" nanoplane. This provides an efficient mechanism for producing dense nanogratings with extremely short periods beyond that allowed by the nanoplasma wave model. In addition, it is also found in our simulation that for an array of nanoplasma sheets of the same parameters mentioned above, an array of intensity maxima as shown in Fig. 3(b) can be produced as well. The simulation result agrees well with our experimental observation in Fig. 2. Our simulation data (not shown here) show that the same phenomenon can be observed for a nanograting consisting of arbitrary number of nanoplanes.

To conclude, we have performed a systematical investigation on the evolution of formation of nanogratings until the grating period reaches a stable value. We reveal an unexpected behavior of splitting of the nanograting at large number of irradiation pulses, which is caused by the interference pattern of the scattering light from the nanograting formed in the early stage. In principle, the splitting process could repeat again and again. However, in reality, it eventually terminates due to the thermal effects, leading to the saturation in the reduction of grating period. The unique mechanism enables formation of nanogratings with periods far beyond that predicted by either the nanoplasmonic or the surface plasma wave model. In combination of our previous observation in Ref. 11, the results provide a more complete physical picture on the formation of nanogratings in bulk glass.

This research was financially supported by National Basic Research Program of China (2014CB921300), and National Science Foundation of China (NSFC) (61275205, 11174305), and the Youth Innovation Promotion Association of Chinese Academy of Sciences.

References
[1] K. Sugioka, and Y. Cheng, Appl. Phys. Rev. 1, 041303 (2014).
[2] S. Kawata, H. B. Sun, T. Tanaka, and K. Takada, Nature 412, 697-698 (2001).
[3] A. P. Joglekar, H. Liu, E. Meyhofer, G. Mourou, and A. J. Hunt, Proc. Natl. Acad. Sci. U. S. A. 101, 5856 (2004).
[4] Y. Liao, Y. Shen, L. Qiao, D. Chen, Y. Cheng, K. Sugioka, and K. Midorikawa, Opt. Lett. 38, 187-189 (2013).
[5] M. Beresna, M. Gecevičius, and P. G. Kazansky, Adv. Opt. Photon. 6 293–339 (2014).
[6] R. Buividas, M. Mikutisc,d, and S. Juodkazisa, Prog. Quantum Electron. 38, 119-156 (2014).
[7] S. Richter, M. Heinrich, S. Doring, A. Tünnermann, S. Nolte, and U. Peschel, J. Laser Appl. 24, 042008 (2012).
[8] R. Taylor, C. Hnatovsky, and E. Simova, Laser Photon. Rev. 2, 26-46 (2008).
[9] Y. Bellouard, M. Dugan, A. A. Said, and P. Bado, Appl. Phys. Lett. 89, 161911-3 (2006).
[10] Y. Bellouard, E. Barthel, A. A. Said, M. Dugan, and P. Bado, Opt. Express 16, 19520-19534 (2008).
[11] Y. Liao, J. Ni, L. Qiao, M. Huang, Y. Bellouard, K. Sugioka, and Y. Cheng, Optica 2, 329-334 (2015).
[12] S. Richter, M. Heinrich, S. Döring, A. Tünnermann, and S. Nolte, Appl. Phys. A 104, 503-507 (2011).
[13] V. R. Bhardwaj, E. Simova, P. P. Rajeev, C. Hnatovsky, R. S. Taylor, D. M. Rayner, and P. B. Corkum, Phys. Rev. Lett. 96, 057404 (2006).
[14] T. H. Elmer, Porous and Reconstructed Glasses. In Engineered Materials Handbook (ASM International, 1992), Volume 4, pp. 427-432.
[15] Y. Cheng, K. Sugioka, K. Midorikawa, M. Masuda, K. Toyoda, M. Kawachi, and K. Shihoyama, Opt. Lett. 28, 55-57 (2003).
[16] G. Obara, H. Shimizu, T. Enami, E. Mazur, M. Terakawa, and M. Obara, Opt. Express 21(22), 26323–26334 (2013).
[17] R. Buschlinger, S. Nolte, and U. Peschel, Phys. Rev. B 89, 184306 (2014).
[18] F. A. Umran, Y. Liao, M. M. Elias, K. Sugioka, R. Stoian, G. Cheng, and Y. Cheng, Opt.Express 21, 15259–15267 (2013).
[19] C. V. Shank, R. Yen, and C. Hirlimann, Phys. Rev. Lett. 50, 454-457 (1983).
[20] K. Sokolowski-Tinten, and D. von der Linde, Phys. Rev. B 61(4), 2643–2650 (2000).